\documentclass[reprint,showpacs,showkeay,preprintnumbers,amsmath,amssymb,aps,prb,floatfix]{revtex4-1}

\usepackage[usenames]{color}
\usepackage{graphicx}
\usepackage{dcolumn}
\usepackage{multirow}
\usepackage{bm}
\usepackage{subfigure}
\usepackage[
  colorlinks=true,
  urlcolor=blue,
  linkcolor=blue,
  citecolor=blue
]{hyperref}

\begin{document}

\title{Electronic and optical properties of spinel zinc
  ferrite:\\ \textit{Ab initio} hybrid functional calculations}

\author{Daniel Fritsch} \affiliation{Department of Chemistry,
  University of Bath, Claverton Down, Bath BA2 7AY, United Kingdom}

\date{\today}

\begin{abstract}
  Spinel ferrites in general show a rich interplay of structural,
  electronic, and magnetic properties. Here, we particularly focus on
  zinc ferrite (ZFO), which has been observed experimentally to
  crystallise in the cubic \textit{normal} spinel structure. However,
  its magnetic ground state is still under dispute. In addition, some
  unusual magnetic properties in ZFO thin films or nanostructures have
  been explained by a possible partial cation inversion and a
  different magnetic interaction between the two cation sublattices of
  the spinel structure compared to the crystalline bulk
  material. Here, density functional theory has been applied to
  investigate the influence of different inversion degrees and
  magnetic couplings among the cation sublattices on the structural,
  electronic, magnetic, and optical properties. Effects of exchange
  and correlation have been modelled using the generalised gradient
  approximation (GGA) together with the Hubbard ``+$U$'' parameter,
  and the more elaborate hybrid functional PBE0. While the GGA+$U$
  calculations yield an antiferromagnetically coupled \textit{normal}
  spinel structure as the ground state, in the PBE0 calculations the
  ferromagnetically coupled \textit{normal} spinel is energetically
  slightly favoured, and the hybrid functional calculations perform
  much better with respect to structural, electronic and optical
  properties.
\end{abstract}

\pacs{71.15.Mb,71.20.Nr,78.20.Bh}

\maketitle

Zinc ferrite or ZnFe$_2$O$_4$ (ZFO) belongs to the mineral class of
spinel ferrites crystallising in a cubic crystal structure with the
general formula $M$Fe$_{2}$O$_{4}$. Common to all spinel ferrites is
the face-centred cubic (fcc) arrangement of the oxygen anions. In the
so-called \textit{normal} spinel structure the octahedrally
coordinated B-sites ($O_{h}$) are solely populated with Fe$^{3+}$
cations whereas the tetrahedrally coordinated A-sites ($T_{d}$)
accomodate a divalent $d$-metal cation, e.g., $M\in$ {Mn, Fe, Co, Ni,
  Cu, Zn, Mg, Cd}.~\cite{SmitWijn_Ferrites} One also finds
mixed-valence occupancies,
($M_{1-\lambda}$Fe$_{\lambda}$)[$M_{\lambda}$Fe$_{2-\lambda}$]O$_{4}$,
with the inversion parameter $\lambda$ ranging from $\lambda=0$ for
the \textit{normal} to $\lambda=1$ for the so-called \textit{inverse}
spinel. Combining a divalent cation with a non-integer inversion
degree, which can partly be tailored by experimental growth
conditions,~\cite{Chen_JPhysD41_205004} offers a huge variety of
structural, electronic, and magnetic properties within the mineral
class of spinel ferrites.

An example for a fully \textit{inverse} spinel ferrite is
NiFe$_2$O$_4$ (NFO), whereas CoFe$_2$O$_4$ (CFO) has a partially
inverse structure with $\lambda \approx 0.8$.~\cite{Brabers1995189}
Both, NFO and CFO, are ferro(i)magnetic (FM) insulators with high
Curie temperatures and large saturation
magnetisations,~\cite{Fritsch_PRB82_104117} with possible applications
in artifical multiferroic heterostructures or spintronics
devices.~\cite{Caffrey_PRB87_024419} An example for a \textit{normal}
spinel is ZFO. Although experimentally mostly reported to be an
antiferromagnetic (AFM) insulator, with a comparatively low N\'{e}el
temperature of $T_{\rm N} \approx$ 10
K,~\cite{Brabers1995189,Schiessl_PRB53_9143,Rabe_AnnuRevCondensMatterPhys1_211}
there is still an ongoing discussion about its magnetic ground
state. While very early models suggested various AFM arrangements of
the magnetic moments among the A- and B-site
sublattices,~\cite{Hastings_PR102_1460,Fayek_PSS37_843,Boucher_PSS40_71,Koenig_SSC8_759}
in recent experiments on high-quality single crystals, no AFM order
has been observed for temperatures down to 1.5 K, and the AFM order is
actually thought to originate from defects or
inhomogeneities.~\cite{Kamazawa_PRB68_024412}

This makes the magnetic structure of spinel ferrites particularly
interesting, e.g. the interactions among the magnetic cations
distributed over the A- and B-sites, and possible phase transitions. A
first understanding of the magnetic interactions in spinels was
provided by the N\'{e}el model~\cite{Neel_AnnPhys3_137} which
described a collinear arrangement of magnetic moments among the A- and
B-site sublattices. This is realised, e.g., in NFO and CFO, where the
A- and B-site cations order ferromagnetically among themselves, and
antiferromagnetically with respect to the other sublattice. With the
$d^{5}$ (Fe$^{3+}$) and $d^{7}$ (Co$^{2+}$) ``levels'' of $d$-orbital
population the additional choice of high-spin vs low-spin complexes
arises. However, for the tetrahedrally coordinated A-sites only
high-spin complexes are observed, and within the mentioned materials
the B-sites have high-spin complexes as well. Different inversion
degrees combined with possible high-spin vs low-spin complexes add
another complication to properly describe the correct magnetic ground
state of spinel ferrites. With a non-magnetic cation (such as
Zn$^{2+}$ in ZFO) populating the A-sites it becomes more favourable
for the Fe$^{3+}$ cations to order antiferromagnetically on the
B-sites below 10 K; contrary to the ferromagnetic coupling in NFO,
CFO, and ZFO at higher
temperatures.~\cite{Brabers1995189,Rabe_AnnuRevCondensMatterPhys1_211}

In addition, contrary to the \textit{normal} spinel ground state of
bulk ZFO, several properties in strained thin films and nanoparticles,
like increased magnetisation, have been explained based on a partial
inversion and/or a AFM to FM change in the B-site
magnetisation. Experimental works so far dealt with single
crystals,~\cite{Kamazawa_PRB68_024412} epitaxial thin
films,~\cite{Bohra_APL88_262506,Chen_JPhysD41_205004,Boentgen_JAP113_073503,Jin_JAP115_213908,LiskovaJakubisova_JAP117_17B726}
and nanocrystalline
samples,~\cite{Ayyappan_APL96_143106,Wu_APL99_202505}
respectively. The optical properties of ZFO have only been rarely
investigated.~\cite{Boentgen_JAP113_073503,LiskovaJakubisova_JAP117_17B726,Zviagin_APL108_131901,Ziaei_EurPhysJB90_29}
Available theoretical works focused on the ground state properties of
ZFO using various exchange-correlation functionals, including variants
of the generalised gradient approximation
(GGA),~\cite{Cheng_PRB78_132403,Soliman_PRB83_083205,RodriguezTorres_PRB89_104414}
different parametrisations of the GGA$+U$
functional,~\cite{Cheng_PRB78_132403,OBrien_JPCM25_445008,Jin_JAP115_213908}
and a recent GGA plus many-body correction
investigation,~\cite{Ziaei_EurPhysJB90_29} respectively. Although more
elaborate hybrid functionals have proven to give improved results for
other spinel ferrites,~\cite{Rowan_PRB79_205103} systematic studies
for ZFO are missing to date.

Here, we focus on differences between the \textit{collinear} AFM and
FM \textit{normal} spinel structure, as well as inverse structures
with $\lambda=0.5$ and $\lambda=1.0$. Presented results include
structural (lattice constant $a$, oxygen parameter $u$, bulk modulus
$B$), electronic (density of states (DOS), band structure), magnetic,
and optical (dielectric functions) properties.

\section{Computational details}

The results of the present work have been obtained by density
functional theory (DFT) calculations employing the Vienna \textit{ab
  initio} Simulation package
(VASP)~\cite{Kresse_PRB47_558,Kresse_PRB49_14251,Kresse_CompMatSci6_15}
together with the projector-augmented wave (PAW)
formalism.~\cite{Bloechl_PRB50_17953} Spin-polarisation has been taken
into account. Standard VASP PAW potentials were used with 14, 12, and
6 valence electrons for iron, zinc, and oxygen,
respectively. Approximations to the exchange-correlation potential
used in the present work are the GGA together with Hubbard ``$+U$''
method~\cite{Anisimov_PRB44_943} based on the Perdew-Burke-Ernzerhof
(PBE) parametrisation,~\cite{Perdew_PRL77_3865} and the more elaborate
hybrid functional PBE0,~\cite{Adamo_JCP110_6158} where a quarter of
the exchange potential is replaced by Hartree-Fock exact-exchange. The
Hubbard ``$+U$'' method has been applied in the simplified,
rotationally invariant version of Dudarev \textit{et
  al.},~\cite{Dudarev_PRB57_1505} with a value of $U_{\rm
  eff}=U-J=5.25$ eV applied to the Fe cations. This value of $U_{\rm
  eff}$ has been chosen to reproduce the experimental band gap of ZFO
of around 2.2
eV,~\cite{Manikandan_JMMM349_249,Peeters_ACSSustChemEng5_2917} is
similar to the $U_{\rm eff}=5.9$ eV applied to
FeO~\cite{Anisimov_PRB44_943} and is slightly larger than $U_{\rm
  eff}=3$ eV applied to NFO and
CFO.~\cite{Fritsch_PRB82_104117,Gutierrez_PRB86_125309} Both
functionals, the GGA$+U$ and the hybrid PBE0, are employed to overcome
the insufficient description of electronic exchange effects in
standard DFT calculations.~\cite{Anisimov_PRB44_943,Adamo_JCP110_6158}

The spinel structure crystallises in the cubic space group
Fd$\bar{3}$m (No. 227) with the smallest possible unit cell
accomodating two functional units (f.u.) of ZFO and 14 atoms. Due to
the necessary application of periodic boundary conditions, the
calculations face an artificial symmetry reduction depending on the
chosen inversion degree and magnetic
order.~\cite{Fritsch_PRB82_104117,Fritsch_APL99_081916,Fritsch_PRB86_014406}
Similar to our earlier works on spinel ferrites, structural
relaxations were performed for several volumes of the smallest
possible unit cell within a scalar-relativistic approximation and
allowing the internal structure parameter to relax until all forces on
the atoms were smaller than 0.001 eV{\AA}$^{-1}$. Dense
$\Gamma$-centred $k$ point meshes of $8 \times 8 \times 8$ for the
Brillouin zone integration and a cutoff energy of 800 eV ensured
converged structural parameters and total energies within meV accuracy
for the GGA$+U$ calculations. For the more demanding PBE0
calculations, $\Gamma$-centred $k$ point meshes of $6 \times 6 \times
6$ and a cutoff energy of 500 eV have been used. Ground-state
geometries were determined by a cubic spline fit to the total energies
with respect to the unit cell volumes.

\section{Results and Discussion}

\begin{table*}
  \caption{\label{Table_StructuralProperties} Ground state structural
    properties (lattice constant $a$, oxygen parameter $u$, and bulk
    modulus $B$) of ZFO calculated with the GGA+$U$ and PBE0
    exchange-correlation functionals and different inversion degrees
    $\lambda$ in comparison to available experimental results. The
    theoretical data also includes the energy difference $\Delta E$
    with respect to the most stable configuration. For the
    \textit{normal} spinel ($\lambda=0.0$) the first (second) data set
    reports structural properties for the AFM (FM) arrangement of
    magnetic moments among the B-site sublattice, respectively.}
  \begin{ruledtabular}
    \begin{tabular}{cccccccccccc}
      
      & \multicolumn{4}{c}{GGA+$U$} & \multicolumn{4}{c}{PBE0} & \multicolumn{3}{c}{Expt.} \\
      
      \cline{2-5} \cline{6-9} \cline{10-12}
      
      $\lambda$ & $a$ & $u$ & $B$ & $\Delta E$ & $a$ & $u$ & $B$ & $\Delta E$ & $a$ & $u$ & $B$ \\
      & [\AA] & & [GPa] & [eV] & [\AA] & & [GPa] & [eV] & [\AA] & & [GPa] \\
      
      0.0 & 8.541 & 0.261 & 161.1 & 0.0 & 8.446 & 0.261 & 173.8 & 0.030 & 8.441~[\onlinecite{Waerenborgh_JSolidStateChem111_300}] & 0.260~[\onlinecite{Hastings_RMP25_114}] & 175.0~[\onlinecite{Reichmann_AmMineral98_601}] \\
      
      & 8.52~[\onlinecite{Cheng_PRB78_132403}] & & & 0.0~[\onlinecite{Cheng_PRB78_132403}] & & & & & 8.4599~[\onlinecite{Schiessl_PRB53_9143}] \\
      
      0.0 & 8.545 & 0.260 & 162.0 & 0.051 & 8.452 & 0.260 & 174.1 & 0.0 \\
      & 8.53~[\onlinecite{Cheng_PRB78_132403}] & & & 0.068~[\onlinecite{Cheng_PRB78_132403}] \\
      
      0.5 & 8.526 & 0.259 & 166.3 & 0.519 & 8.424 & 0.259 & 192.6 & 0.175 & 8.377~[\onlinecite{Wu_APL99_202505}] \\
      
      1.0 & 8.515 & 0.250 & 169.8 & 0.761 & 8.407 & 0.250 & 203.7 & 0.213 \\
      & 8.49~[\onlinecite{Cheng_PRB78_132403}] & & & 0.210~[\onlinecite{Cheng_PRB78_132403}] \\
      
    \end{tabular}
  \end{ruledtabular}
\end{table*}
  
The obtained structural properties of ZFO are given in
table~\ref{Table_StructuralProperties} in comparison to available
experimental results. The GGA$+U$ calculated lattice constants, oxygen
parameters $u$, and bulk moduli for the \textit{normal} spinel are
8.541 {\AA}, 0.261, and 161.1 GPa, and 8.545 {\AA}, 0.260, and 162.0
GPa for the AFM and FM arrangement of magnetic moments among the
B-site sublattice, respectively. The respective PBE0 calculated
lattice constants, oxygen parameters $u$, and bulk moduli are 8.446
{\AA}, 0.261, and 173.8 GPa, and 8.452 {\AA}, 0.260, 174.1 GPa, and
are in much better agreement to the experimental results of 8.441
{\AA}~[\onlinecite{Waerenborgh_JSolidStateChem111_300}] (8.4599
{\AA}~[\onlinecite{Schiessl_PRB53_9143}]),
0.260~[\onlinecite{Hastings_RMP25_114}], and 175.0
GPa~[\onlinecite{Reichmann_AmMineral98_601}], respectively. This is in
line with the general trend that hybrid functional calculations for
oxide semiconductors yield better agreement of lattice parameters and
bulk moduli with respect to experimental
results.~\cite{Fritsch_NRL12_19}

The AFM arrangement of magnetic moments among the B-site sublattice is
the ground state for the GGA+$U$ calculations, in agreement with
experiment and other theoretical investigations, whereas the FM
arrangement is slightly favoured for the hybrid functional PBE0
calculations.

As can be seen from the data in
table~\ref{Table_StructuralProperties}, with increasing inversion
degree $\lambda$ the lattice constants decrease from the
\textit{normal} spinel AFM values for both, GGA$+U$ and PBE0
functional calculations, in agreement with experimental results of Wu
\textit{et al.},~\cite{Wu_APL99_202505} who report a lattice constant
of 8.377 {\AA} for a partially inverse ZFO sample with
$\lambda=0.5$. Both, the GGA$+U$ and PBE0 calculated bulk moduli
increase with increasing inversion degree $\lambda$.

\begin{figure}[t]
  \begin{center}
    \includegraphics[width=\columnwidth,clip]{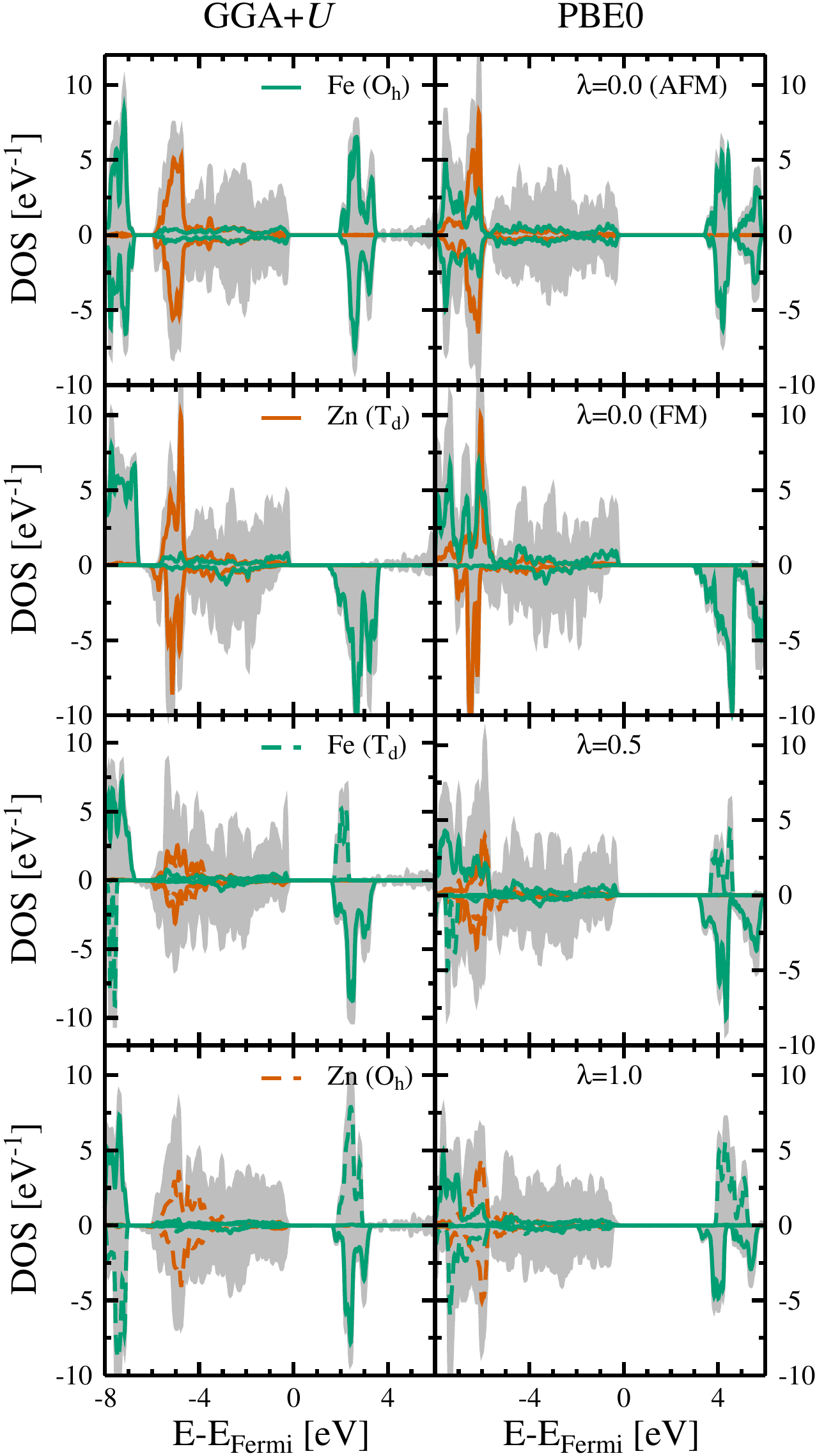}
    \caption{\label{Fig_DOS}(Color online) Total and projected DOS per
      formula unit for ZFO calculated with the GGA$+U$ (left panels)
      and PBE0 (right panels) exchange-correlation functionals for
      different inversion degrees $\lambda$ and arrangements of the
      magnetic moments among the B-site sublattice, respectively. The
      octahedral $O_{h}$ (tetrahedral $T_{d}$) states of Fe are shown
      as straight (dashed) green lines, and the tetrahedral $T_{d}$
      (octahedral $O_{h}$) states of Zn are shown as straight (dashed)
      red lines, respectively. The shaded grey area in all panels
      depicts the total DOS. Minority spin projections are shown using
      negative values. The zero energy is set to the valence band
      maximum.}
  \end{center}
\end{figure}

The DOS of \textit{normal} ZFO with a AFM and FM arrangement of
magnetic moments among the B-site sublattices, partially inverse ZFO
($\lambda=0.5$) and fully inverse ZFO ($\lambda=1.0$) are shown in
figure~\ref{Fig_DOS}. The left and right panels depict the results of
GGA+$U$ and PBE0 calculations, and the zero energy is set to the
valence band maximum, respectively. In each of the panels the total
DOS is shown as shaded grey area, and the spin-up (spin-down)
contributions are given along the positive (negative)
$y$-axis. Generally, apart from the larger band gap for the PBE0
calculations, the overall shape of the DOS looks quite similar for the
GGA$+U$ and the PBE0 calculations.

From the DOS with respect to the inversion degree $\lambda$ it can be
seen that an increasing inversion degree shifts the valence band DOS
towards higher energies, and exhibits a growing contribution to the
tetrahedral Fe and octahedral Zn DOS, respectively. From
figure~\ref{Fig_DOS} it can also be seen that both, GGA+$U$ and PBE0
calculations, yield an insulating behaviour, underlying the important
effect of electronic exchange effects on the electronic properties
compared to plain GGA
approaches.~\cite{Fritsch_JPhysConfSer292_012014}

The local magnetic moments for Fe cations are found to be in the range
of 4.1 to 4.3 $\mu_{B}$, with the GGA$+U$ values being slightly larger
than the PBE0 ones, for all inversion degrees and arrangements of
magnetic moments among the B-site sublattices. This is in favourable
agreement with the experimental value of 4.22
$\mu_{B}$~[\onlinecite{Schiessl_PRB53_9143}] and the range of 4.1 to
4.2 $\mu_{B}$ reported from other GGA$+U$
calculations,~\cite{Cheng_PRB78_132403} respectively.

\begin{figure}[t]
  \begin{center} 
    \includegraphics[width=\columnwidth,clip]{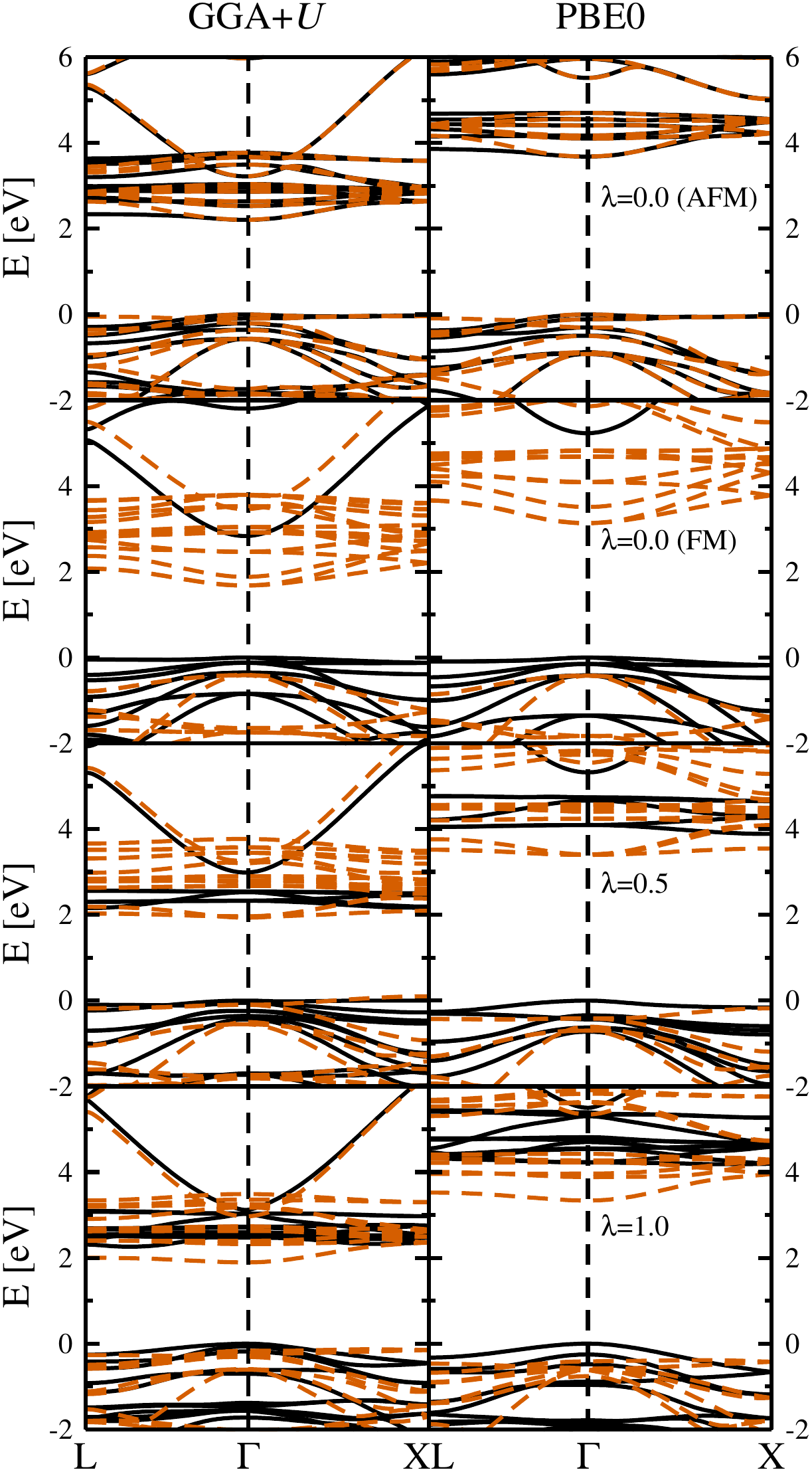}
    \caption{\label{Fig_BAND}(Color online) Electronic band structures
      for ZFO calculated with the GGA$+U$ (left panels) and PBE0
      (right panels) exchange-correlation functionals for different
      inversion degrees $\lambda$ and arrangements of the magnetic
      moments among the B-site sublattice, respectively. Majority- and
      minority-spin bands are shown as full (black) and dashed (red)
      lines. The zero energy is set to the valence band maximum.}
  \end{center}
\end{figure}

For the \textit{normal} spinel and AFM arrangement of magnetic moments
among the B-site sublattice it is no surprise that the spin resolved
band structure depicted in the upper panels of figure~\ref{Fig_BAND}
shows he same behaviour for the spin-up and spin-down channels, with a
band gap of 2.21 eV (3.68 eV) for the GGA$+U$ (PBE0)
calculations. While the GGA$+U$ band gap is close to the experimental
band gap of around 2.2
eV,~\cite{Manikandan_JMMM349_249,Peeters_ACSSustChemEng5_2917} the
PBE0 band gap is close to the experimental band gap of 3.31 eV,
obtained from spectroscopic ellipsometry
measurements.~\cite{Boentgen_JAP113_073503} This changes drastically
for the FM arrangement, where the band gap is now between the spin-up
valence band and the spin-down conduction band for both, and amounts
to 1.68 eV (3.13 eV) for the GGA+$U$ (PBE0) calculations,
respectively. Changing the inversion degree $\lambda$ from 0.5 to 1.0
only has a small influence on the band gap, which is now around 1.91
eV (3.37 eV) for the GGA$+U$ (PBE0) calculations.

\begin{figure}[t]
  \begin{center}
    \includegraphics[width=\columnwidth,clip]{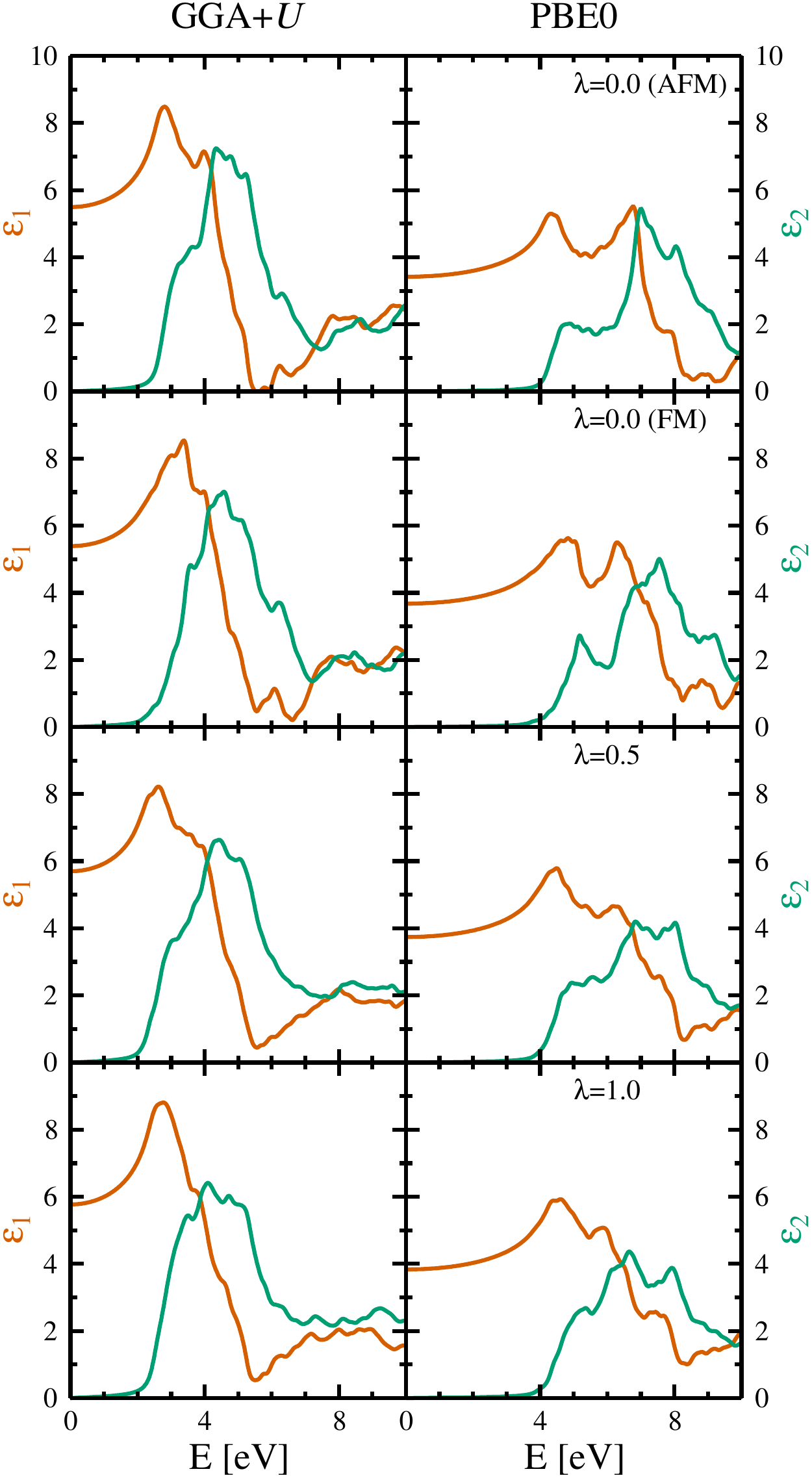}
    \caption{\label{Fig_DielFunc}(Color online) Real (red) and
      imaginary (green) part of the dielectric functions for ZFO
      calculated with the GGA$+U$ (left panels) and PBE0 (right
      panels) exchange-correlation functionals for different inversion
      degrees $\lambda$ and arrangements of the magnetic moments among
      the B-site sublattice, respectively.}
  \end{center}
\end{figure}

Based on the obtained relaxed structural ground states for different
inversion degrees $\lambda$ and magnetic arrangements on the B-site
sublattices the optical properties have been calculated and are
depicted in figure~\ref{Fig_DielFunc}. Generally, the onsets of the
imaginary parts of the PBE0 dielectric functions ($\varepsilon_{2}$)
are shifted towards higher energies, as to be expected from the larger
band gaps already seen in the DOS (figure~\ref{Fig_DOS}) and the band
structures (figure~\ref{Fig_BAND}). Moreover, all the PBE0 dielectric
functions are flatter compared to the GGA$+U$ ones. If one accounts
for the slight changes in the observed band gaps, the overall features
of the imaginary part of the PBE0 dielectric function for
\textit{normal} ZFO and a AFM arrangement of the magnetic moment among
the B-site cations (upper right panel of figure~\ref{Fig_DielFunc}),
is in good agreement with the optical data reported
in.~\cite{Boentgen_JAP113_073503}

\section{Summary and Outlook}

In summary, a detailed investigation on the influence of the inversion
degree $\lambda$ and different arrangements of the magnetic moments
among the B-site cations on the structural, electronic, magnetic, and
optical properties of the spinel ferrite ZFO has been presented. It
has been shown that the hybrid functional PBE0 performs better than
GGA$+U$, and yields an overall better agreement with respect to
experimental results on structural, electronic and optical
properties. These hybrid functional calculations provide the basis for
future in-depth analyses of the optical properties in relation to the
band structure and DOS, as well as taking into account strain effects
from underlying
substrates.~\cite{Fritsch_PRB82_104117,Fritsch_PRB86_014406}

\acknowledgements

This research has received funding from the European Union’s Horizon
2020 research and innovation programme under grant agreement No 641864
(INREP). This work made use of the ARCHER UK National Supercomputing
Service (http://www.archer.ac.uk) via the membership of the UK’s HPC
Materials Chemistry Consortium, funded by EPSRC (EP/L000202) and the
Balena HPC facility of the University of Bath.


\begin{thebibliography}{44}%
\makeatletter
\providecommand \@ifxundefined [1]{%
 \@ifx{#1\undefined}
}%
\providecommand \@ifnum [1]{%
 \ifnum #1\expandafter \@firstoftwo
 \else \expandafter \@secondoftwo
 \fi
}%
\providecommand \@ifx [1]{%
 \ifx #1\expandafter \@firstoftwo
 \else \expandafter \@secondoftwo
 \fi
}%
\providecommand \natexlab [1]{#1}%
\providecommand \enquote  [1]{``#1''}%
\providecommand \bibnamefont  [1]{#1}%
\providecommand \bibfnamefont [1]{#1}%
\providecommand \citenamefont [1]{#1}%
\providecommand \href@noop [0]{\@secondoftwo}%
\providecommand \href [0]{\begingroup \@sanitize@url \@href}%
\providecommand \@href[1]{\@@startlink{#1}\@@href}%
\providecommand \@@href[1]{\endgroup#1\@@endlink}%
\providecommand \@sanitize@url [0]{\catcode `\\12\catcode `\$12\catcode
  `\&12\catcode `\#12\catcode `\^12\catcode `\_12\catcode `\%12\relax}%
\providecommand \@@startlink[1]{}%
\providecommand \@@endlink[0]{}%
\providecommand \url  [0]{\begingroup\@sanitize@url \@url }%
\providecommand \@url [1]{\endgroup\@href {#1}{\urlprefix }}%
\providecommand \urlprefix  [0]{URL }%
\providecommand \Eprint [0]{\href }%
\providecommand \doibase [0]{http://dx.doi.org/}%
\providecommand \selectlanguage [0]{\@gobble}%
\providecommand \bibinfo  [0]{\@secondoftwo}%
\providecommand \bibfield  [0]{\@secondoftwo}%
\providecommand \translation [1]{[#1]}%
\providecommand \BibitemOpen [0]{}%
\providecommand \bibitemStop [0]{}%
\providecommand \bibitemNoStop [0]{.\EOS\space}%
\providecommand \EOS [0]{\spacefactor3000\relax}%
\providecommand \BibitemShut  [1]{\csname bibitem#1\endcsname}%
\let\auto@bib@innerbib\@empty
\bibitem [{\citenamefont {Smit}\ and\ \citenamefont
  {Wijn}(1959)}]{SmitWijn_Ferrites}%
  \BibitemOpen
  \bibfield  {author} {\bibinfo {author} {\bibfnamefont {J.}~\bibnamefont
  {Smit}}\ and\ \bibinfo {author} {\bibfnamefont {H.~P.~J.}\ \bibnamefont
  {Wijn}},\ }\href@noop {} {\emph {\bibinfo {title} {Ferrites}}}\ (\bibinfo
  {publisher} {Cleaver-Hume Press},\ \bibinfo {address} {London},\ \bibinfo
  {year} {1959})\BibitemShut {NoStop}%
\bibitem [{\citenamefont {Chen}\ \emph {et~al.}(2008)\citenamefont {Chen},
  \citenamefont {Spoddig},\ and\ \citenamefont {Ziese}}]{Chen_JPhysD41_205004}%
  \BibitemOpen
  \bibfield  {author} {\bibinfo {author} {\bibfnamefont {Y.~F.}\ \bibnamefont
  {Chen}}, \bibinfo {author} {\bibfnamefont {D.}~\bibnamefont {Spoddig}}, \
  and\ \bibinfo {author} {\bibfnamefont {M.}~\bibnamefont {Ziese}},\ }\href
  {\doibase 10.1088/0022-3727/41/20/205004} {\bibfield  {journal} {\bibinfo
  {journal} {J. Phys. D: Appl. Phys.}\ }\textbf {\bibinfo {volume} {41}},\
  \bibinfo {pages} {205004} (\bibinfo {year} {2008})}\BibitemShut {NoStop}%
\bibitem [{\citenamefont {Brabers}(1995)}]{Brabers1995189}%
  \BibitemOpen
  \bibfield  {author} {\bibinfo {author} {\bibfnamefont {V.~A.~M.}\
  \bibnamefont {Brabers}},\ }in\ \href {\doibase 10.1016/S1567-2719(05)80032-0}
  {\emph {\bibinfo {booktitle} {Handbook of Magnetic Materials}}},\
  Vol.~\bibinfo {volume} {8},\ \bibinfo {editor} {edited by\ \bibinfo {editor}
  {\bibfnamefont {K.~H.~J.}\ \bibnamefont {Buschow}}}\ (\bibinfo  {publisher}
  {Elsevier},\ \bibinfo {year} {1995})\ pp.\ \bibinfo {pages} {189 --
  324}\BibitemShut {NoStop}%
\bibitem [{\citenamefont {Fritsch}\ and\ \citenamefont
  {Ederer}(2010)}]{Fritsch_PRB82_104117}%
  \BibitemOpen
  \bibfield  {author} {\bibinfo {author} {\bibfnamefont {D.}~\bibnamefont
  {Fritsch}}\ and\ \bibinfo {author} {\bibfnamefont {C.}~\bibnamefont
  {Ederer}},\ }\href {\doibase 10.1103/PhysRevB.82.104117} {\bibfield
  {journal} {\bibinfo  {journal} {Phys. Rev. B}\ }\textbf {\bibinfo {volume}
  {82}},\ \bibinfo {pages} {104117} (\bibinfo {year} {2010})}\BibitemShut
  {NoStop}%
\bibitem [{\citenamefont {Caffrey}\ \emph {et~al.}(2013)\citenamefont
  {Caffrey}, \citenamefont {Fritsch}, \citenamefont {Archer}, \citenamefont
  {Sanvito},\ and\ \citenamefont {Ederer}}]{Caffrey_PRB87_024419}%
  \BibitemOpen
  \bibfield  {author} {\bibinfo {author} {\bibfnamefont {N.}~\bibnamefont
  {Caffrey}}, \bibinfo {author} {\bibfnamefont {D.}~\bibnamefont {Fritsch}},
  \bibinfo {author} {\bibfnamefont {T.}~\bibnamefont {Archer}}, \bibinfo
  {author} {\bibfnamefont {S.}~\bibnamefont {Sanvito}}, \ and\ \bibinfo
  {author} {\bibfnamefont {C.}~\bibnamefont {Ederer}},\ }\href {\doibase
  10.1103/PhysRevB.87.024419} {\bibfield  {journal} {\bibinfo  {journal} {Phys.
  Rev. B}\ }\textbf {\bibinfo {volume} {87}},\ \bibinfo {pages} {024419}
  (\bibinfo {year} {2013})}\BibitemShut {NoStop}%
\bibitem [{\citenamefont {Schiessl}\ \emph {et~al.}(1996)\citenamefont
  {Schiessl}, \citenamefont {Potzel}, \citenamefont {Karzel}, \citenamefont
  {Steiner}, \citenamefont {Kalvius}, \citenamefont {Martin}, \citenamefont
  {Krause}, \citenamefont {Halevy}, \citenamefont {Gal}, \citenamefont
  {Sch\"{a}fer}, \citenamefont {Will}, \citenamefont {Hillberg},\ and\
  \citenamefont {W\"{a}ppling}}]{Schiessl_PRB53_9143}%
  \BibitemOpen
  \bibfield  {author} {\bibinfo {author} {\bibfnamefont {W.}~\bibnamefont
  {Schiessl}}, \bibinfo {author} {\bibfnamefont {W.}~\bibnamefont {Potzel}},
  \bibinfo {author} {\bibfnamefont {H.}~\bibnamefont {Karzel}}, \bibinfo
  {author} {\bibfnamefont {M.}~\bibnamefont {Steiner}}, \bibinfo {author}
  {\bibfnamefont {G.~M.}\ \bibnamefont {Kalvius}}, \bibinfo {author}
  {\bibfnamefont {A.}~\bibnamefont {Martin}}, \bibinfo {author} {\bibfnamefont
  {M.~K.}\ \bibnamefont {Krause}}, \bibinfo {author} {\bibfnamefont
  {I.}~\bibnamefont {Halevy}}, \bibinfo {author} {\bibfnamefont
  {J.}~\bibnamefont {Gal}}, \bibinfo {author} {\bibfnamefont {W.}~\bibnamefont
  {Sch\"{a}fer}}, \bibinfo {author} {\bibfnamefont {G.}~\bibnamefont {Will}},
  \bibinfo {author} {\bibfnamefont {M.}~\bibnamefont {Hillberg}}, \ and\
  \bibinfo {author} {\bibfnamefont {R.}~\bibnamefont {W\"{a}ppling}},\ }\href
  {\doibase 10.1103/PhysRevB.53.9143} {\bibfield  {journal} {\bibinfo
  {journal} {Phys. Rev. B}\ }\textbf {\bibinfo {volume} {53}},\ \bibinfo
  {pages} {9143} (\bibinfo {year} {1996})}\BibitemShut {NoStop}%
\bibitem [{\citenamefont {Rabe}(2010)}]{Rabe_AnnuRevCondensMatterPhys1_211}%
  \BibitemOpen
  \bibfield  {author} {\bibinfo {author} {\bibfnamefont {K.~M.}\ \bibnamefont
  {Rabe}},\ }\href {\doibase 10.1146/annurev-conmatphys-070909-103932}
  {\bibfield  {journal} {\bibinfo  {journal} {Annu. Rev. Condens. Matter
  Phys.}\ }\textbf {\bibinfo {volume} {1}},\ \bibinfo {pages} {211} (\bibinfo
  {year} {2010})}\BibitemShut {NoStop}%
\bibitem [{\citenamefont {Hastings}\ and\ \citenamefont
  {Corliss}(1956)}]{Hastings_PR102_1460}%
  \BibitemOpen
  \bibfield  {author} {\bibinfo {author} {\bibfnamefont {J.~M.}\ \bibnamefont
  {Hastings}}\ and\ \bibinfo {author} {\bibfnamefont {L.~M.}\ \bibnamefont
  {Corliss}},\ }\href {\doibase 10.1103/PhysRev.102.1460} {\bibfield  {journal}
  {\bibinfo  {journal} {Phys. Rev.}\ }\textbf {\bibinfo {volume} {102}},\
  \bibinfo {pages} {1460} (\bibinfo {year} {1956})}\BibitemShut {NoStop}%
\bibitem [{\citenamefont {Fayek}\ \emph {et~al.}(1970)\citenamefont {Fayek},
  \citenamefont {Leciejewicz}, \citenamefont {Murasik},\ and\ \citenamefont
  {Yamzin}}]{Fayek_PSS37_843}%
  \BibitemOpen
  \bibfield  {author} {\bibinfo {author} {\bibfnamefont {M.~K.}\ \bibnamefont
  {Fayek}}, \bibinfo {author} {\bibfnamefont {J.}~\bibnamefont {Leciejewicz}},
  \bibinfo {author} {\bibfnamefont {A.}~\bibnamefont {Murasik}}, \ and\
  \bibinfo {author} {\bibfnamefont {I.~I.}\ \bibnamefont {Yamzin}},\ }\href
  {\doibase 10.1002/pssb.19700370235} {\bibfield  {journal} {\bibinfo
  {journal} {physica status solidi}\ }\textbf {\bibinfo {volume} {37}},\
  \bibinfo {pages} {843} (\bibinfo {year} {1970})}\BibitemShut {NoStop}%
\bibitem [{\citenamefont {Boucher}\ \emph {et~al.}(1970)\citenamefont
  {Boucher}, \citenamefont {Buhl},\ and\ \citenamefont
  {Perrin}}]{Boucher_PSS40_71}%
  \BibitemOpen
  \bibfield  {author} {\bibinfo {author} {\bibfnamefont {B.}~\bibnamefont
  {Boucher}}, \bibinfo {author} {\bibfnamefont {R.}~\bibnamefont {Buhl}}, \
  and\ \bibinfo {author} {\bibfnamefont {M.}~\bibnamefont {Perrin}},\ }\href
  {\doibase 10.1002/pssb.19700400118} {\bibfield  {journal} {\bibinfo
  {journal} {physica status solidi}\ }\textbf {\bibinfo {volume} {40}},\
  \bibinfo {pages} {171} (\bibinfo {year} {1970})}\BibitemShut {NoStop}%
\bibitem [{\citenamefont {K\"onig}\ \emph {et~al.}(1970)\citenamefont
  {K\"onig}, \citenamefont {Bertaut}, \citenamefont {Gros}, \citenamefont
  {Mitrikov},\ and\ \citenamefont {Chol}}]{Koenig_SSC8_759}%
  \BibitemOpen
  \bibfield  {author} {\bibinfo {author} {\bibfnamefont {U.}~\bibnamefont
  {K\"onig}}, \bibinfo {author} {\bibfnamefont {E.~F.}\ \bibnamefont
  {Bertaut}}, \bibinfo {author} {\bibfnamefont {Y.}~\bibnamefont {Gros}},
  \bibinfo {author} {\bibfnamefont {M.}~\bibnamefont {Mitrikov}}, \ and\
  \bibinfo {author} {\bibfnamefont {G.}~\bibnamefont {Chol}},\ }\href {\doibase
  10.1016/0038-1098(70)90425-4} {\bibfield  {journal} {\bibinfo  {journal}
  {Solid State Communications}\ }\textbf {\bibinfo {volume} {8}},\ \bibinfo
  {pages} {759} (\bibinfo {year} {1970})}\BibitemShut {NoStop}%
\bibitem [{\citenamefont {Kamazawa}\ \emph {et~al.}(2003)\citenamefont
  {Kamazawa}, \citenamefont {Tsunoda}, \citenamefont {Kadowaki},\ and\
  \citenamefont {Kohn}}]{Kamazawa_PRB68_024412}%
  \BibitemOpen
  \bibfield  {author} {\bibinfo {author} {\bibfnamefont {K.}~\bibnamefont
  {Kamazawa}}, \bibinfo {author} {\bibfnamefont {Y.}~\bibnamefont {Tsunoda}},
  \bibinfo {author} {\bibfnamefont {H.}~\bibnamefont {Kadowaki}}, \ and\
  \bibinfo {author} {\bibfnamefont {K.}~\bibnamefont {Kohn}},\ }\href {\doibase
  10.1103/PhysRevB.68.024412} {\bibfield  {journal} {\bibinfo  {journal} {Phys.
  Rev. B}\ }\textbf {\bibinfo {volume} {68}},\ \bibinfo {pages} {024412}
  (\bibinfo {year} {2003})}\BibitemShut {NoStop}%
\bibitem [{\citenamefont {N\'{e}el}(1948)}]{Neel_AnnPhys3_137}%
  \BibitemOpen
  \bibfield  {author} {\bibinfo {author} {\bibfnamefont {L.}~\bibnamefont
  {N\'{e}el}},\ }\href@noop {} {\bibfield  {journal} {\bibinfo  {journal} {Ann.
  Phys. (Paris)}\ }\textbf {\bibinfo {volume} {3}},\ \bibinfo {pages} {137}
  (\bibinfo {year} {1948})}\BibitemShut {NoStop}%
\bibitem [{\citenamefont {Bohra}\ \emph {et~al.}(2006)\citenamefont {Bohra},
  \citenamefont {Prasada}, \citenamefont {Kumar}, \citenamefont {Misra},\ and\
  \citenamefont {Sahoo}}]{Bohra_APL88_262506}%
  \BibitemOpen
  \bibfield  {author} {\bibinfo {author} {\bibfnamefont {M.}~\bibnamefont
  {Bohra}}, \bibinfo {author} {\bibfnamefont {S.}~\bibnamefont {Prasada}},
  \bibinfo {author} {\bibfnamefont {N.}~\bibnamefont {Kumar}}, \bibinfo
  {author} {\bibfnamefont {D.~S.}\ \bibnamefont {Misra}}, \ and\ \bibinfo
  {author} {\bibfnamefont {S.~C.}\ \bibnamefont {Sahoo}},\ }\href {\doibase
  10.1063/1.2217253} {\bibfield  {journal} {\bibinfo  {journal} {Appl. Phys.
  Lett.}\ }\textbf {\bibinfo {volume} {88}},\ \bibinfo {pages} {262506}
  (\bibinfo {year} {2006})}\BibitemShut {NoStop}%
\bibitem [{\citenamefont {B\"{o}ntgen}\ \emph {et~al.}(2013)\citenamefont
  {B\"{o}ntgen}, \citenamefont {Brachwitz}, \citenamefont {Schmidt-Grund},
  \citenamefont {Lorenz},\ and\ \citenamefont
  {Grundmann}}]{Boentgen_JAP113_073503}%
  \BibitemOpen
  \bibfield  {author} {\bibinfo {author} {\bibfnamefont {T.}~\bibnamefont
  {B\"{o}ntgen}}, \bibinfo {author} {\bibfnamefont {K.}~\bibnamefont
  {Brachwitz}}, \bibinfo {author} {\bibfnamefont {R.}~\bibnamefont
  {Schmidt-Grund}}, \bibinfo {author} {\bibfnamefont {M.}~\bibnamefont
  {Lorenz}}, \ and\ \bibinfo {author} {\bibfnamefont {M.}~\bibnamefont
  {Grundmann}},\ }\href {\doibase 10.1063/1.4790881} {\bibfield  {journal}
  {\bibinfo  {journal} {J. Appl. Phys.}\ }\textbf {\bibinfo {volume} {113}},\
  \bibinfo {pages} {073503} (\bibinfo {year} {2013})}\BibitemShut {NoStop}%
\bibitem [{\citenamefont {Jin}\ \emph {et~al.}(2014)\citenamefont {Jin},
  \citenamefont {Li}, \citenamefont {Mi},\ and\ \citenamefont
  {Bai}}]{Jin_JAP115_213908}%
  \BibitemOpen
  \bibfield  {author} {\bibinfo {author} {\bibfnamefont {C.}~\bibnamefont
  {Jin}}, \bibinfo {author} {\bibfnamefont {P.}~\bibnamefont {Li}}, \bibinfo
  {author} {\bibfnamefont {W.}~\bibnamefont {Mi}}, \ and\ \bibinfo {author}
  {\bibfnamefont {H.}~\bibnamefont {Bai}},\ }\href {\doibase 10.1063/1.4881502}
  {\bibfield  {journal} {\bibinfo  {journal} {J. Appl. Phys.}\ }\textbf
  {\bibinfo {volume} {115}},\ \bibinfo {pages} {213908} (\bibinfo {year}
  {2014})}\BibitemShut {NoStop}%
\bibitem [{\citenamefont {Li\u{s}kov\'{a}-Jakubisov\'{a}}\ \emph
  {et~al.}(2015)\citenamefont {Li\u{s}kov\'{a}-Jakubisov\'{a}}, \citenamefont
  {\u{S}. Vi\u{s}\u{n}ovsk\'{y}}, \citenamefont {\u{S}irok\'{y}}, \citenamefont
  {Hrabovsk\'{y}}, \citenamefont {Pi\u{s}tora}, \citenamefont {Sahoo},
  \citenamefont {Prasad}, \citenamefont {Venkataramani}, \citenamefont
  {Bohra},\ and\ \citenamefont {Krishnan}}]{LiskovaJakubisova_JAP117_17B726}%
  \BibitemOpen
  \bibfield  {author} {\bibinfo {author} {\bibfnamefont {E.}~\bibnamefont
  {Li\u{s}kov\'{a}-Jakubisov\'{a}}}, \bibinfo {author} {\bibnamefont {\u{S}.
  Vi\u{s}\u{n}ovsk\'{y}}}, \bibinfo {author} {\bibfnamefont {P.}~\bibnamefont
  {\u{S}irok\'{y}}}, \bibinfo {author} {\bibfnamefont {D.}~\bibnamefont
  {Hrabovsk\'{y}}}, \bibinfo {author} {\bibfnamefont {J.}~\bibnamefont
  {Pi\u{s}tora}}, \bibinfo {author} {\bibfnamefont {S.~C.}\ \bibnamefont
  {Sahoo}}, \bibinfo {author} {\bibfnamefont {S.}~\bibnamefont {Prasad}},
  \bibinfo {author} {\bibfnamefont {N.}~\bibnamefont {Venkataramani}}, \bibinfo
  {author} {\bibfnamefont {M.}~\bibnamefont {Bohra}}, \ and\ \bibinfo {author}
  {\bibfnamefont {R.}~\bibnamefont {Krishnan}},\ }\href {\doibase
  10.1063/1.4916936} {\bibfield  {journal} {\bibinfo  {journal} {J. Appl.
  Phys.}\ }\textbf {\bibinfo {volume} {117}},\ \bibinfo {pages} {17B726}
  (\bibinfo {year} {2015})}\BibitemShut {NoStop}%
\bibitem [{\citenamefont {Ayyappan}\ \emph {et~al.}(2010)\citenamefont
  {Ayyappan}, \citenamefont {Raja}, \citenamefont {Venkateswaran},
  \citenamefont {Philip},\ and\ \citenamefont {Raj}}]{Ayyappan_APL96_143106}%
  \BibitemOpen
  \bibfield  {author} {\bibinfo {author} {\bibfnamefont {S.}~\bibnamefont
  {Ayyappan}}, \bibinfo {author} {\bibfnamefont {S.~P.}\ \bibnamefont {Raja}},
  \bibinfo {author} {\bibfnamefont {C.}~\bibnamefont {Venkateswaran}}, \bibinfo
  {author} {\bibfnamefont {J.}~\bibnamefont {Philip}}, \ and\ \bibinfo {author}
  {\bibfnamefont {B.}~\bibnamefont {Raj}},\ }\href {\doibase 10.1063/1.3374332}
  {\bibfield  {journal} {\bibinfo  {journal} {Appl. Phys. Lett.}\ }\textbf
  {\bibinfo {volume} {96}},\ \bibinfo {pages} {143106} (\bibinfo {year}
  {2010})}\BibitemShut {NoStop}%
\bibitem [{\citenamefont {Wu}\ \emph {et~al.}(2011)\citenamefont {Wu},
  \citenamefont {Li}, \citenamefont {Xu}, \citenamefont {Jiang}, \citenamefont
  {Ye}, \citenamefont {Xie},\ and\ \citenamefont {Zheng}}]{Wu_APL99_202505}%
  \BibitemOpen
  \bibfield  {author} {\bibinfo {author} {\bibfnamefont {J.}~\bibnamefont
  {Wu}}, \bibinfo {author} {\bibfnamefont {N.}~\bibnamefont {Li}}, \bibinfo
  {author} {\bibfnamefont {J.}~\bibnamefont {Xu}}, \bibinfo {author}
  {\bibfnamefont {Y.}~\bibnamefont {Jiang}}, \bibinfo {author} {\bibfnamefont
  {Z.-G.}\ \bibnamefont {Ye}}, \bibinfo {author} {\bibfnamefont
  {Z.}~\bibnamefont {Xie}}, \ and\ \bibinfo {author} {\bibfnamefont
  {L.}~\bibnamefont {Zheng}},\ }\href {\doibase 10.1063/1.3662840} {\bibfield
  {journal} {\bibinfo  {journal} {Appl. Phys. Lett.}\ }\textbf {\bibinfo
  {volume} {99}},\ \bibinfo {pages} {202505} (\bibinfo {year}
  {2011})}\BibitemShut {NoStop}%
\bibitem [{\citenamefont {Zviagin}\ \emph {et~al.}(2016)\citenamefont
  {Zviagin}, \citenamefont {Kumar}, \citenamefont {Lorite}, \citenamefont
  {Esquinazi}, \citenamefont {Grundmann},\ and\ \citenamefont
  {Schmidt-Grund}}]{Zviagin_APL108_131901}%
  \BibitemOpen
  \bibfield  {author} {\bibinfo {author} {\bibfnamefont {V.}~\bibnamefont
  {Zviagin}}, \bibinfo {author} {\bibfnamefont {Y.}~\bibnamefont {Kumar}},
  \bibinfo {author} {\bibfnamefont {I.}~\bibnamefont {Lorite}}, \bibinfo
  {author} {\bibfnamefont {P.}~\bibnamefont {Esquinazi}}, \bibinfo {author}
  {\bibfnamefont {M.}~\bibnamefont {Grundmann}}, \ and\ \bibinfo {author}
  {\bibfnamefont {R.}~\bibnamefont {Schmidt-Grund}},\ }\href {\doibase
  10.1063/1.4944898} {\bibfield  {journal} {\bibinfo  {journal} {Appl. Phys.
  Lett.}\ }\textbf {\bibinfo {volume} {108}},\ \bibinfo {pages} {131901}
  (\bibinfo {year} {2016})}\BibitemShut {NoStop}%
\bibitem [{\citenamefont {Ziaei}\ and\ \citenamefont
  {Bredow}(2017)}]{Ziaei_EurPhysJB90_29}%
  \BibitemOpen
  \bibfield  {author} {\bibinfo {author} {\bibfnamefont {V.}~\bibnamefont
  {Ziaei}}\ and\ \bibinfo {author} {\bibfnamefont {T.}~\bibnamefont {Bredow}},\
  }\href {\doibase 10.1140/epjb/e2017-70623-9} {\bibfield  {journal} {\bibinfo
  {journal} {Eur. Phys. J. B}\ }\textbf {\bibinfo {volume} {90}},\ \bibinfo
  {pages} {29} (\bibinfo {year} {2017})}\BibitemShut {NoStop}%
\bibitem [{\citenamefont {Cheng}(2008)}]{Cheng_PRB78_132403}%
  \BibitemOpen
  \bibfield  {author} {\bibinfo {author} {\bibfnamefont {C.}~\bibnamefont
  {Cheng}},\ }\href {\doibase 10.1103/PhysRevB.78.132403} {\bibfield  {journal}
  {\bibinfo  {journal} {Phys. Rev. B}\ }\textbf {\bibinfo {volume} {78}},\
  \bibinfo {pages} {132403} (\bibinfo {year} {2008})}\BibitemShut {NoStop}%
\bibitem [{\citenamefont {Soliman}\ \emph {et~al.}(2011)\citenamefont
  {Soliman}, \citenamefont {Elfalaky}, \citenamefont {Fecher},\ and\
  \citenamefont {Felser}}]{Soliman_PRB83_083205}%
  \BibitemOpen
  \bibfield  {author} {\bibinfo {author} {\bibfnamefont {S.}~\bibnamefont
  {Soliman}}, \bibinfo {author} {\bibfnamefont {A.}~\bibnamefont {Elfalaky}},
  \bibinfo {author} {\bibfnamefont {G.~H.}\ \bibnamefont {Fecher}}, \ and\
  \bibinfo {author} {\bibfnamefont {C.}~\bibnamefont {Felser}},\ }\href
  {\doibase 10.1103/PhysRevB.83.085205} {\bibfield  {journal} {\bibinfo
  {journal} {Phys. Rev. B}\ }\textbf {\bibinfo {volume} {83}},\ \bibinfo
  {pages} {083205} (\bibinfo {year} {2011})}\BibitemShut {NoStop}%
\bibitem [{\citenamefont {Torres}\ \emph {et~al.}(2014)\citenamefont {Torres},
  \citenamefont {Pasquevich}, \citenamefont {Z\'elis}, \citenamefont {Golmar},
  \citenamefont {Heluani}, \citenamefont {Nayak}, \citenamefont {Adeagbo},
  \citenamefont {Hergert}, \citenamefont {Hoffmann}, \citenamefont {Ernst},
  \citenamefont {Esquinazi},\ and\ \citenamefont
  {Stewart}}]{RodriguezTorres_PRB89_104414}%
  \BibitemOpen
  \bibfield  {author} {\bibinfo {author} {\bibfnamefont {C.~E.~R.}\
  \bibnamefont {Torres}}, \bibinfo {author} {\bibfnamefont {G.~A.}\
  \bibnamefont {Pasquevich}}, \bibinfo {author} {\bibfnamefont {P.~M.}\
  \bibnamefont {Z\'elis}}, \bibinfo {author} {\bibfnamefont {F.}~\bibnamefont
  {Golmar}}, \bibinfo {author} {\bibfnamefont {S.~P.}\ \bibnamefont {Heluani}},
  \bibinfo {author} {\bibfnamefont {S.~K.}\ \bibnamefont {Nayak}}, \bibinfo
  {author} {\bibfnamefont {W.~A.}\ \bibnamefont {Adeagbo}}, \bibinfo {author}
  {\bibfnamefont {W.}~\bibnamefont {Hergert}}, \bibinfo {author} {\bibfnamefont
  {M.}~\bibnamefont {Hoffmann}}, \bibinfo {author} {\bibfnamefont
  {A.}~\bibnamefont {Ernst}}, \bibinfo {author} {\bibfnamefont
  {P.}~\bibnamefont {Esquinazi}}, \ and\ \bibinfo {author} {\bibfnamefont
  {S.~J.}\ \bibnamefont {Stewart}},\ }\href {\doibase
  10.1103/PhysRevB.89.104411} {\bibfield  {journal} {\bibinfo  {journal} {Phys.
  Rev. B}\ }\textbf {\bibinfo {volume} {89}},\ \bibinfo {pages} {104414}
  (\bibinfo {year} {2014})}\BibitemShut {NoStop}%
\bibitem [{\citenamefont {O'Brien}\ \emph {et~al.}(2013)\citenamefont
  {O'Brien}, \citenamefont {R\'ak},\ and\ \citenamefont
  {Brenner}}]{OBrien_JPCM25_445008}%
  \BibitemOpen
  \bibfield  {author} {\bibinfo {author} {\bibfnamefont {C.~J.}\ \bibnamefont
  {O'Brien}}, \bibinfo {author} {\bibfnamefont {Z.}~\bibnamefont {R\'ak}}, \
  and\ \bibinfo {author} {\bibfnamefont {D.~W.}\ \bibnamefont {Brenner}},\
  }\href {\doibase 10.1088/0953-8984/25/44/445008} {\bibfield  {journal}
  {\bibinfo  {journal} {J. Phys.: Condens. Matter}\ }\textbf {\bibinfo {volume}
  {25}},\ \bibinfo {pages} {445008} (\bibinfo {year} {2013})}\BibitemShut
  {NoStop}%
\bibitem [{\citenamefont {Rowan}\ \emph {et~al.}(2009)\citenamefont {Rowan},
  \citenamefont {Patterson},\ and\ \citenamefont
  {Gasparov}}]{Rowan_PRB79_205103}%
  \BibitemOpen
  \bibfield  {author} {\bibinfo {author} {\bibfnamefont {A.~D.}\ \bibnamefont
  {Rowan}}, \bibinfo {author} {\bibfnamefont {C.~H.}\ \bibnamefont
  {Patterson}}, \ and\ \bibinfo {author} {\bibfnamefont {L.~V.}\ \bibnamefont
  {Gasparov}},\ }\href {\doibase 10.1103/PhysRevB.79.205103} {\bibfield
  {journal} {\bibinfo  {journal} {Phys. Rev. B}\ }\textbf {\bibinfo {volume}
  {79}},\ \bibinfo {pages} {205103} (\bibinfo {year} {2009})}\BibitemShut
  {NoStop}%
\bibitem [{\citenamefont {Kresse}\ and\ \citenamefont
  {Hafner}(1993)}]{Kresse_PRB47_558}%
  \BibitemOpen
  \bibfield  {author} {\bibinfo {author} {\bibfnamefont {G.}~\bibnamefont
  {Kresse}}\ and\ \bibinfo {author} {\bibfnamefont {J.}~\bibnamefont
  {Hafner}},\ }\href {\doibase 10.1103/PhysRevB.47.558} {\bibfield  {journal}
  {\bibinfo  {journal} {Phys. Rev. B}\ }\textbf {\bibinfo {volume} {47}},\
  \bibinfo {pages} {558} (\bibinfo {year} {1993})}\BibitemShut {NoStop}%
\bibitem [{\citenamefont {Kresse}\ and\ \citenamefont
  {Hafner}(1994)}]{Kresse_PRB49_14251}%
  \BibitemOpen
  \bibfield  {author} {\bibinfo {author} {\bibfnamefont {G.}~\bibnamefont
  {Kresse}}\ and\ \bibinfo {author} {\bibfnamefont {J.}~\bibnamefont
  {Hafner}},\ }\href {\doibase 10.1103/PhysRevB.49.14251} {\bibfield  {journal}
  {\bibinfo  {journal} {Phys. Rev. B}\ }\textbf {\bibinfo {volume} {49}},\
  \bibinfo {pages} {14251} (\bibinfo {year} {1994})}\BibitemShut {NoStop}%
\bibitem [{\citenamefont {Kresse}\ and\ \citenamefont
  {Furthm\"uller}(1996)}]{Kresse_CompMatSci6_15}%
  \BibitemOpen
  \bibfield  {author} {\bibinfo {author} {\bibfnamefont {G.}~\bibnamefont
  {Kresse}}\ and\ \bibinfo {author} {\bibfnamefont {J.}~\bibnamefont
  {Furthm\"uller}},\ }\href {\doibase 10.1016/0927-0256(96)00008-0} {\bibfield
  {journal} {\bibinfo  {journal} {Comp. Mat. Sci.}\ }\textbf {\bibinfo {volume}
  {6}},\ \bibinfo {pages} {15} (\bibinfo {year} {1996})}\BibitemShut {NoStop}%
\bibitem [{\citenamefont {Bl\"ochl}(1994)}]{Bloechl_PRB50_17953}%
  \BibitemOpen
  \bibfield  {author} {\bibinfo {author} {\bibfnamefont {P.~E.}\ \bibnamefont
  {Bl\"ochl}},\ }\href {\doibase 10.1103/PhysRevB.50.17953} {\bibfield
  {journal} {\bibinfo  {journal} {Phys. Rev. B}\ }\textbf {\bibinfo {volume}
  {50}},\ \bibinfo {pages} {17953} (\bibinfo {year} {1994})}\BibitemShut
  {NoStop}%
\bibitem [{\citenamefont {Anisimov}\ \emph {et~al.}(1991)\citenamefont
  {Anisimov}, \citenamefont {Zaanen},\ and\ \citenamefont
  {Andersen}}]{Anisimov_PRB44_943}%
  \BibitemOpen
  \bibfield  {author} {\bibinfo {author} {\bibfnamefont {V.~I.}\ \bibnamefont
  {Anisimov}}, \bibinfo {author} {\bibfnamefont {J.}~\bibnamefont {Zaanen}}, \
  and\ \bibinfo {author} {\bibfnamefont {O.~K.}\ \bibnamefont {Andersen}},\
  }\href {\doibase 10.1103/PhysRevB.44.943} {\bibfield  {journal} {\bibinfo
  {journal} {Phys. Rev. B}\ }\textbf {\bibinfo {volume} {44}},\ \bibinfo
  {pages} {943} (\bibinfo {year} {1991})}\BibitemShut {NoStop}%
\bibitem [{\citenamefont {Perdew}\ \emph {et~al.}(1996)\citenamefont {Perdew},
  \citenamefont {Burke},\ and\ \citenamefont {Ernzerhof}}]{Perdew_PRL77_3865}%
  \BibitemOpen
  \bibfield  {author} {\bibinfo {author} {\bibfnamefont {J.~P.}\ \bibnamefont
  {Perdew}}, \bibinfo {author} {\bibfnamefont {K.}~\bibnamefont {Burke}}, \
  and\ \bibinfo {author} {\bibfnamefont {M.}~\bibnamefont {Ernzerhof}},\ }\href
  {\doibase 10.1103/PhysRevLett.77.3865} {\bibfield  {journal} {\bibinfo
  {journal} {Phys. Rev. Lett.}\ }\textbf {\bibinfo {volume} {77}},\ \bibinfo
  {pages} {3865} (\bibinfo {year} {1996})}\BibitemShut {NoStop}%
\bibitem [{\citenamefont {Adamo}\ and\ \citenamefont
  {Barone}(1999)}]{Adamo_JCP110_6158}%
  \BibitemOpen
  \bibfield  {author} {\bibinfo {author} {\bibfnamefont {C.}~\bibnamefont
  {Adamo}}\ and\ \bibinfo {author} {\bibfnamefont {V.}~\bibnamefont {Barone}},\
  }\href {\doibase 10.1063/1.478522} {\bibfield  {journal} {\bibinfo  {journal}
  {J. Chem. Phys.}\ }\textbf {\bibinfo {volume} {110}},\ \bibinfo {pages}
  {6158} (\bibinfo {year} {1999})}\BibitemShut {NoStop}%
\bibitem [{\citenamefont {Dudarev}\ \emph {et~al.}(1998)\citenamefont
  {Dudarev}, \citenamefont {Botton}, \citenamefont {Savrasov}, \citenamefont
  {Humphreys},\ and\ \citenamefont {Sutton}}]{Dudarev_PRB57_1505}%
  \BibitemOpen
  \bibfield  {author} {\bibinfo {author} {\bibfnamefont {S.~L.}\ \bibnamefont
  {Dudarev}}, \bibinfo {author} {\bibfnamefont {G.~A.}\ \bibnamefont {Botton}},
  \bibinfo {author} {\bibfnamefont {S.~Y.}\ \bibnamefont {Savrasov}}, \bibinfo
  {author} {\bibfnamefont {C.~J.}\ \bibnamefont {Humphreys}}, \ and\ \bibinfo
  {author} {\bibfnamefont {A.~P.}\ \bibnamefont {Sutton}},\ }\href {\doibase
  10.1103/PhysRevB.57.1505} {\bibfield  {journal} {\bibinfo  {journal} {Phys.
  Rev. B}\ }\textbf {\bibinfo {volume} {57}},\ \bibinfo {pages} {1505}
  (\bibinfo {year} {1998})}\BibitemShut {NoStop}%
\bibitem [{\citenamefont {Manikandan}\ \emph {et~al.}(2014)\citenamefont
  {Manikandan}, \citenamefont {Kennedy}, \citenamefont {Bououdina},\ and\
  \citenamefont {Vijaya}}]{Manikandan_JMMM349_249}%
  \BibitemOpen
  \bibfield  {author} {\bibinfo {author} {\bibfnamefont {A.}~\bibnamefont
  {Manikandan}}, \bibinfo {author} {\bibfnamefont {L.~J.}\ \bibnamefont
  {Kennedy}}, \bibinfo {author} {\bibfnamefont {M.}~\bibnamefont {Bououdina}},
  \ and\ \bibinfo {author} {\bibfnamefont {J.~J.}\ \bibnamefont {Vijaya}},\
  }\href {\doibase 10.1016/j.jmmm.2013.09.013} {\bibfield  {journal} {\bibinfo
  {journal} {J. Magn. Magn. Mat.}\ }\textbf {\bibinfo {volume} {349}},\
  \bibinfo {pages} {249} (\bibinfo {year} {2014})}\BibitemShut {NoStop}%
\bibitem [{\citenamefont {Peeters}\ \emph {et~al.}(2017)\citenamefont
  {Peeters}, \citenamefont {Taffa}, \citenamefont {Kerrigan}, \citenamefont
  {Ney}, \citenamefont {J\"{o}ns}, \citenamefont {Rogalla}, \citenamefont
  {Cwik}, \citenamefont {Becker}, \citenamefont {Grafen}, \citenamefont
  {Ostendorf}, \citenamefont {Winter}, \citenamefont {Chakraborty},
  \citenamefont {Wark},\ and\ \citenamefont
  {Devi}}]{Peeters_ACSSustChemEng5_2917}%
  \BibitemOpen
  \bibfield  {author} {\bibinfo {author} {\bibfnamefont {D.}~\bibnamefont
  {Peeters}}, \bibinfo {author} {\bibfnamefont {D.~H.}\ \bibnamefont {Taffa}},
  \bibinfo {author} {\bibfnamefont {M.~M.}\ \bibnamefont {Kerrigan}}, \bibinfo
  {author} {\bibfnamefont {A.}~\bibnamefont {Ney}}, \bibinfo {author}
  {\bibfnamefont {N.}~\bibnamefont {J\"{o}ns}}, \bibinfo {author}
  {\bibfnamefont {D.}~\bibnamefont {Rogalla}}, \bibinfo {author} {\bibfnamefont
  {S.}~\bibnamefont {Cwik}}, \bibinfo {author} {\bibfnamefont {H.-W.}\
  \bibnamefont {Becker}}, \bibinfo {author} {\bibfnamefont {M.}~\bibnamefont
  {Grafen}}, \bibinfo {author} {\bibfnamefont {A.}~\bibnamefont {Ostendorf}},
  \bibinfo {author} {\bibfnamefont {C.~H.}\ \bibnamefont {Winter}}, \bibinfo
  {author} {\bibfnamefont {S.}~\bibnamefont {Chakraborty}}, \bibinfo {author}
  {\bibfnamefont {M.}~\bibnamefont {Wark}}, \ and\ \bibinfo {author}
  {\bibfnamefont {A.}~\bibnamefont {Devi}},\ }\href {\doibase
  10.1021/acssuschemeng.6b02233} {\bibfield  {journal} {\bibinfo  {journal}
  {ACS Sustainable Chem. Eng.}\ }\textbf {\bibinfo {volume} {5}},\ \bibinfo
  {pages} {2917} (\bibinfo {year} {2017})}\BibitemShut {NoStop}%
\bibitem [{\citenamefont {Guti{\'e}rrez}\ \emph {et~al.}(2012)\citenamefont
  {Guti{\'e}rrez}, \citenamefont {Foerster}, \citenamefont {Fina},
  \citenamefont {Fontcuberta}, \citenamefont {Fritsch},\ and\ \citenamefont
  {Ederer}}]{Gutierrez_PRB86_125309}%
  \BibitemOpen
  \bibfield  {author} {\bibinfo {author} {\bibfnamefont {D.}~\bibnamefont
  {Guti{\'e}rrez}}, \bibinfo {author} {\bibfnamefont {M.}~\bibnamefont
  {Foerster}}, \bibinfo {author} {\bibfnamefont {I.}~\bibnamefont {Fina}},
  \bibinfo {author} {\bibfnamefont {J.}~\bibnamefont {Fontcuberta}}, \bibinfo
  {author} {\bibfnamefont {D.}~\bibnamefont {Fritsch}}, \ and\ \bibinfo
  {author} {\bibfnamefont {C.}~\bibnamefont {Ederer}},\ }\href {\doibase
  10.1103/PhysRevB.86.125309} {\bibfield  {journal} {\bibinfo  {journal} {Phys.
  Rev. B}\ }\textbf {\bibinfo {volume} {86}},\ \bibinfo {pages} {125309}
  (\bibinfo {year} {2012})}\BibitemShut {NoStop}%
\bibitem [{\citenamefont {Fritsch}\ and\ \citenamefont
  {Ederer}(2011{\natexlab{a}})}]{Fritsch_APL99_081916}%
  \BibitemOpen
  \bibfield  {author} {\bibinfo {author} {\bibfnamefont {D.}~\bibnamefont
  {Fritsch}}\ and\ \bibinfo {author} {\bibfnamefont {C.}~\bibnamefont
  {Ederer}},\ }\href {\doibase 10.1063/1.3631676} {\bibfield  {journal}
  {\bibinfo  {journal} {Appl. Phys. Lett.}\ }\textbf {\bibinfo {volume} {99}},\
  \bibinfo {pages} {081916} (\bibinfo {year} {2011}{\natexlab{a}})}\BibitemShut
  {NoStop}%
\bibitem [{\citenamefont {Fritsch}\ and\ \citenamefont
  {Ederer}(2012)}]{Fritsch_PRB86_014406}%
  \BibitemOpen
  \bibfield  {author} {\bibinfo {author} {\bibfnamefont {D.}~\bibnamefont
  {Fritsch}}\ and\ \bibinfo {author} {\bibfnamefont {C.}~\bibnamefont
  {Ederer}},\ }\href {\doibase 10.1103/PhysRevB.86.014406} {\bibfield
  {journal} {\bibinfo  {journal} {Phys. Rev. B}\ }\textbf {\bibinfo {volume}
  {86}},\ \bibinfo {pages} {014406} (\bibinfo {year} {2012})}\BibitemShut
  {NoStop}%
\bibitem [{\citenamefont {Waerenborgh}\ \emph {et~al.}(1994)\citenamefont
  {Waerenborgh}, \citenamefont {Figueiredo}, \citenamefont {Cabral},\ and\
  \citenamefont {Pereira}}]{Waerenborgh_JSolidStateChem111_300}%
  \BibitemOpen
  \bibfield  {author} {\bibinfo {author} {\bibfnamefont {J.~C.}\ \bibnamefont
  {Waerenborgh}}, \bibinfo {author} {\bibfnamefont {M.~O.}\ \bibnamefont
  {Figueiredo}}, \bibinfo {author} {\bibfnamefont {J.~M.~P.}\ \bibnamefont
  {Cabral}}, \ and\ \bibinfo {author} {\bibfnamefont {L.~C.~J.}\ \bibnamefont
  {Pereira}},\ }\href {\doibase 10.1006/jssc.1994.1231} {\bibfield  {journal}
  {\bibinfo  {journal} {J. Solid State Chem.}\ }\textbf {\bibinfo {volume}
  {111}},\ \bibinfo {pages} {300} (\bibinfo {year} {1994})}\BibitemShut
  {NoStop}%
\bibitem [{\citenamefont {Hastings}\ and\ \citenamefont
  {Corliss}(1953)}]{Hastings_RMP25_114}%
  \BibitemOpen
  \bibfield  {author} {\bibinfo {author} {\bibfnamefont {J.~M.}\ \bibnamefont
  {Hastings}}\ and\ \bibinfo {author} {\bibfnamefont {L.~M.}\ \bibnamefont
  {Corliss}},\ }\href {\doibase 10.1103/RevModPhys.25.114} {\bibfield
  {journal} {\bibinfo  {journal} {Rev. Mod. Phys.}\ }\textbf {\bibinfo {volume}
  {25}},\ \bibinfo {pages} {114} (\bibinfo {year} {1953})}\BibitemShut
  {NoStop}%
\bibitem [{\citenamefont {Reichmann}\ \emph {et~al.}(2013)\citenamefont
  {Reichmann}, \citenamefont {Jacobsen},\ and\ \citenamefont
  {Ballaran}}]{Reichmann_AmMineral98_601}%
  \BibitemOpen
  \bibfield  {author} {\bibinfo {author} {\bibfnamefont {H.~J.}\ \bibnamefont
  {Reichmann}}, \bibinfo {author} {\bibfnamefont {S.~D.}\ \bibnamefont
  {Jacobsen}}, \ and\ \bibinfo {author} {\bibfnamefont {T.~B.}\ \bibnamefont
  {Ballaran}},\ }\href {\doibase 10.2138/am.2013.4294} {\bibfield  {journal}
  {\bibinfo  {journal} {Am. Mineral.}\ }\textbf {\bibinfo {volume} {98}},\
  \bibinfo {pages} {601} (\bibinfo {year} {2013})}\BibitemShut {NoStop}%
\bibitem [{\citenamefont {Fritsch}\ \emph {et~al.}(2017)\citenamefont
  {Fritsch}, \citenamefont {Morgan},\ and\ \citenamefont
  {Walsh}}]{Fritsch_NRL12_19}%
  \BibitemOpen
  \bibfield  {author} {\bibinfo {author} {\bibfnamefont {D.}~\bibnamefont
  {Fritsch}}, \bibinfo {author} {\bibfnamefont {B.~J.}\ \bibnamefont {Morgan}},
  \ and\ \bibinfo {author} {\bibfnamefont {A.}~\bibnamefont {Walsh}},\ }\href
  {\doibase 10.1186/s11671-016-1779-9} {\bibfield  {journal} {\bibinfo
  {journal} {Nanoscale Research Letters}\ }\textbf {\bibinfo {volume} {12}},\
  \bibinfo {pages} {19} (\bibinfo {year} {2017})}\BibitemShut {NoStop}%
\bibitem [{\citenamefont {Fritsch}\ and\ \citenamefont
  {Ederer}(2011{\natexlab{b}})}]{Fritsch_JPhysConfSer292_012014}%
  \BibitemOpen
  \bibfield  {author} {\bibinfo {author} {\bibfnamefont {D.}~\bibnamefont
  {Fritsch}}\ and\ \bibinfo {author} {\bibfnamefont {C.}~\bibnamefont
  {Ederer}},\ }\href {\doibase 10.1088/1742-6596/292/1/012014} {\bibfield
  {journal} {\bibinfo  {journal} {J. Phys.: Conf. Ser.}\ }\textbf {\bibinfo
  {volume} {292}},\ \bibinfo {pages} {012014} (\bibinfo {year}
  {2011}{\natexlab{b}})}\BibitemShut {NoStop}%
\end{thebibliography}
\end{document}